\documentclass[aps,prl,twocolumn,groupedaddress,showpacs,showkeyst]{revtex4-1}

\usepackage{amsmath,amssymb}
\usepackage{graphicx}
\usepackage[usenames,dvipsnames]{xcolor}
\usepackage{enumitem}

\newcommand{\bt}[1]{{\mathbf #1}}
\newcommand{\deltaS}{\delta_{\scalebox{.3}{\emph{S}}}}
\newcommand{\deltaT}{\delta_{\scalebox{.3}{\emph{T}}}}
\newcommand{\deltaSb}{\delta_{\scalebox{.6}{\emph{S}}}}
\newcommand{\deltaTb}{\delta_{\scalebox{.6}{\emph{T}}}}

\newcommand{\mrm}[1]{\mathrm{#1}}

\begin{document}
\title{Robust energy transfer mechanism and critically balanced turbulence\\ via non-resonant triads in nonlinear wave systems}
\author{Miguel D. Bustamante}
\email[]{miguel.bustamante@ucd.ie}
\affiliation{School of Mathematical Sciences,
University College Dublin,
Belfield, Dublin 4, Ireland}
\author{Brenda Quinn}
\email[]{brenda.quinn@ucd.ie}
\affiliation{School of Mathematical Sciences,
University College Dublin,
Belfield, Dublin 4, Ireland}

\date{\today}

\begin{abstract}
A robust energy transfer mechanism is found in nonlinear wave systems, which favours transfers towards modes interacting via non-resonant triads, applicable in meteorology, nonlinear optics and plasma wave turbulence. Transfer efficiency is maximal when the frequency mismatch of the non-resonant triad balances the system's nonlinear frequency: at intermediate levels of oscillation amplitudes an instability is triggered that explores unstable manifolds of periodic orbits, so turbulent cascades are most efficient at intermediate nonlinearity. Numerical simulations confirm analytical predictions.
\end{abstract}

\maketitle

\bibliographystyle{apsrev4-1}

\noindent \textbf{Introduction.} A variety of physical systems of high technological importance consist of nonlinearly interacting oscillations or waves including nonlinear circuits in electrical power systems, high-intensity lasers, nonlinear photonics, gravity water waves in oceans, Rossby-Haurwitz planetary waves in the atmosphere, drift waves in fusion plasmas, etc. These systems are characterised by extreme events that are localised in space and time and are associated with strong nonlinear energy exchanges that dramatically alter the system's global behaviour.
One of the few consistent theories that deal with these nonlinear exchanges is classical wave turbulence theory~\cite{Zakharov1992,Nazarenko2011}. This theory makes ad-hoc assumptions on correlations of the evolving quantities and produces statistical predictions.  One example where this theory is widely used is in numerical prediction of ocean waves.
Despite the success of the theory however, there remains a lack of understanding of the actual physical mechanisms that are responsible for the energy transfers.

This Letter addresses the basic energy transfer mechanisms in real physical systems, precisely in the context where the hypotheses of classical wave turbulence theory do not hold, namely when the spatial domains have a finite size, when the amplitudes of the carrying fields are not infinitesimally small and when the linear wave timescales are comparable to the timescales of the nonlinear oscillations. Once these basic mechanisms are understood, the impact on the field will be multidisciplinary as a new door will open for more accurate models, in line with the ideas of chaos and ergodicity, stable/unstable manifolds, Lyapunov exponents, bifurcations, etc~\cite{Eckmann1985,Cvitanovic91,Waleffe97}.
This new understanding will lead turbulence theory to quantitative predictions which could be confirmed experimentally and numerically~\cite{Newell2011}.

The appropriate framework is the governing partial differential equations (PDE) of classical turbulence, nonlinear optics, quantum fluids and magneto-hydrodynamics considered on bounded physical domains. The corresponding wave field $\Psi(\mathbf{x},t)$ can be decomposed as a sum of Fourier harmonics, $\Psi(\mathbf{x},t) = \sum_{\mathbf{k}} b_{\mathbf{k}}(t) \mathrm{e}^{\mathrm{i} (\mathbf{k}.\mathbf{x} - \omega(\mathbf{k}) t)},$ over a suitable discrete domain e.g. $\mathbb{Z}^d$ and where $\omega(\mathbf{k})$ is the linear dispersion relation. If the PDE is nonlinear, the spectral modes $b_{\mathbf{k}}(t)$ interact and exchange energy amongst themselves as the wave field evolves. These modes are a set of complex functions of time that, depending on the degree of nonlinearity in the PDE, tend to interact in triads (quadratic) or quartets (cubic).  A triad is a group of three spectral modes $b_1(t), b_2(t), b_3(t)$ whose wavevectors $\mathbf{k}_1, \mathbf{k}_2, \mathbf{k}_3$ satisfy 
\begin{eqnarray}
\label{k_resonance}
\mathbf{k}_1+ \mathbf{k}_2 &=& \mathbf{k}_3\,.
\end{eqnarray}
The triad is called resonant if $\omega(\mathbf{k}_1)+\omega(\mathbf{k}_2)=\omega(\mathbf{k}_3)\,.$ Otherwise it is called non-resonant.

Since any mode belongs to several triads, energy can be transferred nonlinearly throughout the intricate network or cluster of connected triads. The theory that deals with these energy exchanges is Discrete and Mesoscopic Wave Turbulence~\cite{Korotkevich2005,Nazarenko2006,Kartashova2007b,Bustamante2009a,Bustamante2009b,Lvov2010,Kartashova2011,Harris2012,Harris2013,Harper2012,Bustamante2013} and is still in development. As evidence for this Letter's timeliness, it was established recently that (i) non-resonant triads/quartets are responsible for most of the energy exchanges in real systems~\cite{Janssen2003,Smith2005,Annenkov2006} and (ii) if a ``critical balance'' between nonlinear and linear frequencies is assumed phenomenologically, an energy spectrum is obtained that matches the predictions of strong-wave turbulence theories \cite{Goldreich1995,Nazarenko2011a}.\\

\noindent \textbf{New Turbulence Paradigm.} For any initial condition in spectral space, we aim to elucidate the basic question of how energy is transferred through the scales. Although the answer to this question in general might seem very difficult because of the nonintegrability of the system of evolving modes, it is possible to achieve a deep understanding of the transfer mechanisms which lead to energy cascades and provides a new paradigm of turbulence.

This Letter's results apply to a variety of systems, including two quadratic PDE models for:\\
(1) {\bf Drift waves}
in inhomogeneous plasmas with wavevector $\bt{k}=(k_x,k_y)$ and $\omega_{\bt k}=\frac{-\beta k_x}{k^2+{1}/{\rho^2}}$ supported by the Hasegawa-Mima equation,
where the wave field is the electrostatic potential, $\rho$ is the ion Larmor radius at the electron temperature
and $\beta$ is a constant proportional to the mean plasma density gradient.\\
(2) {\bf Rossby-Haurwitz waves}
on a sphere which are critical for the distribution of energy in the atmosphere~\cite{Lynch2009}, supported by the barotropic vorticity equation with wavevector $\bt{k}=(m,n)$ and $\omega_{\bt k}=\frac{-2m\Omega_E}{n(n+1)}$ where $\Omega_E$ is the angular velocity of the sphere.

We discuss drift waves from here on. The equation
 governing the evolution of the amplitudes is
\begin{eqnarray}
\label{db_dt}
\partial_t b_{\bt k} &=& \sum\limits_{\textbf{k}_1, \textbf{k}_2 \in \mathbb{Z}^2} Z^{\bt k}_{12}\, \delta^{\bt k}_{12}\, b_{\bt{k}_1}\, b_{\bt{k}_2}\,\mathrm{e}^{\mathrm{i}\,(\omega_{{\bt k}} - \omega_{1} - \omega_{2}) t}.
\vspace{-1.5mm}
\end{eqnarray}
Coefficients $Z^{\bt k}_{12}$ are the interaction coefficients, $\omega_{j}  \equiv \omega({\bt k}_j)$ and the Kronecker symbol $\delta^{\bt k}_{12}$ defines the set of three wave vectors which satisfy Eq.~\eqref{k_resonance}.
  The most generic and robust energy transfer will occur via two-common-mode triad interactions since if any two modes $ \bt{k}_1,  \bt{k}_2$ have non-zero amplitude then the mode $\bt{k}$
will immediately start changing its amplitude because of the nonlinear term in Eq.~\eqref{db_dt}.

The crucial observation is that the nonlinear frequency $\Gamma$ of the oscillations of $b_{{\bt k}_j}$ is typically proportional to the amplitudes of the oscillations. It follows that at some intermediate value of the amplitudes, the nonlinear frequency $\Gamma$ becomes comparable to the linear frequency mismatch $\delta \equiv \omega_{1} + \omega_{2} -\omega_{{\bt k}}.$ Thus, the amplitudes can be re-scaled so that the sum in Eq.~(\ref{db_dt}) contains at least some non-oscillatory terms. These terms lead to sustained growth of the mode $b_{{\bt k}},$ even from zero initial condition. This phenomenon is thus a new nonlinear instability. Its key aspects are:\\
\noindent $\bullet$ For any non-zero initial condition of the source modes, one can re-scale the initial conditions by a common numerical scale factor in order to trigger the new instability towards a chosen target triad.\\
\noindent $\bullet$ This instability is robust with respect to the choice of physical wave system: if, as is customary in turbulence theory, one discards dissipation and forcing terms in the governing equations, then the theory applies.\\
\noindent $\bullet$ For this transfer to be efficient a necessary condition is that the target triad's frequency mismatch be non-zero. In other words, non-resonant triads are capable of receiving, via nonlinear transfers, substantially more energy than exactly resonant triads.\\
These aspects make the new instability truly robust as compared with e.g. decay/modulational instability. It is quite easy to find the effect in numerical simulations and it should be relatively easy to find in experiments. We will now detail the mechanism and later extend the idea to describe turbulence cascades as energy transfers throughout the network of triads.\\

\noindent \textbf{Detailed proof of the instability.} The simplest model illustrating this effect is a two-triad cluster. Four evolving modes with complex amplitudes $b_1(t)$, $b_2(t)$, $b_3(t)$ and $b_4(t),$ correspond to wavenumbers $\mathbf{k}_1, \mathbf{k}_2, \mathbf{k}_3, \mathbf{k}_4$ satisfying the $3$-wave conditions $\mathbf{k}_1+\mathbf{k}_2=\mathbf{k}_3$ and $\mathbf{k}_2+\mathbf{k}_3=\mathbf{k}_4$. Define frequency mismatches  $\delta_S = \omega(\mathbf{k}_1)+\omega(\mathbf{k}_2)-\omega(\mathbf{k}_3),$  $\delta_T = \omega(\mathbf{k}_2)+\omega(\mathbf{k}_3)-\omega(\mathbf{k}_4).$ The cluster evolves according to the system
\begin{eqnarray}
\nonumber
\dot{b}_1 &=& S_1 \, b_2^* b_3 \, \mathrm{e}^{\mathrm{i}\,\deltaS\, t} \\
\nonumber
\dot{b}_2 &=& S_2\, b_1^* b_3 \, \mathrm{e}^{\mathrm{i}\,\deltaS\, t}
+ \epsilon \,T_1\, b_3^* b_4 \, \mathrm{e}^{\mathrm{i}\,\deltaT\, t} \\
\nonumber
\dot{b}_3 &=&  S_3\, b_1 b_2 \, \mathrm{e}^{-\mathrm{i}\,\deltaS\, t}
+ \epsilon \,T_2\, b_2^* b_4 \, \mathrm{e}^{\mathrm{i}\,\deltaT\, t}\\
\label{eq:ODEs}
\dot{b}_4 &=&  T_3\, b_2 b_3 \, \mathrm{e}^{-\mathrm{i}\,\deltaT\, t}.
\vspace{-4.mm}
\end{eqnarray}
The six interaction coefficients $S_j$, $T_j$ are assumed to be real and nonzero. The parameter $\epsilon$ controls the strength of the interaction between the two triads. Only models with bounded evolution are considered here. This implies that $S_1, S_2, S_3$ do not have the same sign, and likewise for $T_1, T_2, T_3$. The full dynamical system (\ref{eq:ODEs}) has two invariants:
\begin{eqnarray}
\nonumber I_1 &=& - S_2 T_3 |b_1|^2 + S_1 T_3 |b_2|^2 - \epsilon\, S_1 T_1 |b_4|^2\\
\label{I1I2} I_2 &=& - S_3 T_3 |b_1|^2 + S_1 T_3 |b_3|^2 - \epsilon\,S_1 T_2 |b_4|^2\,.
\end{eqnarray}
Suppose all energy is initially in the source triad ``S'' i.e., initially $b_4$ is zero but $(b_1, b_2, b_3)$ is nonzero.
The new result consists of an instability in the form of a strong energy transfer from the source triad to the target triad ``T''.  The proof is by contradiction: assume $b_4$ remains small for subsequent times. Then the terms involving $\epsilon$ in system~(\ref{eq:ODEs}) can be neglected
and at this level of approximation, the first three equations form a closed system that describe an isolated triad and can be integrated analytically. For simplicity the case $\delta_{\scalebox{.5}{\emph{S}}} = 0$ is discussed. Three conservation laws are available in the isolated triad: Hamiltonian ${\mathcal{H}} \equiv \mathrm{Im}(b_1 b_2 b_3^*)$ and the Manley-Rowe invariants (\ref{I1I2}) with $\epsilon$-terms discarded~\cite{Craik1988,Lynch2004,Kartashova2011}.

The analytical solution for the isolated triad contains a periodic part and a precession part, and both are important: \mbox{$b_j(t) = B_j(t) \,\mrm{e}^{\mrm{i}\Omega_j\,t}\,,$} where $B_j(t)$ are periodic complex functions \cite{Kartashova2011} with frequency \mbox{$\Gamma = \Gamma(I_1,I_2,{\mathcal{H}}),$} the so-called nonlinear frequency broadening. The nonlinear ``precession frequencies'' $\Omega_j (= \Omega_j(I_1,I_2,{\mathcal{H}}))$ satisfy $\Omega_1 + \Omega_2 = \Omega_3$ and are incommensurate to $\Gamma.$ Explicit formulae for $\Gamma$ and $\Omega_j$ are available as supplemental material. The last Eq.~in (\ref{eq:ODEs}) is integrated by quadratures:
 \begin{eqnarray}
\label{eq:solb4INST}
b_4(t)  &=& \int_0^t F(t') \mrm{e}^{\mrm{i}(\Omega_2+\Omega_3-\deltaT)t'}dt' \,,
\end{eqnarray}
where $F(t) \equiv T_3 \,  B_2(t) B_3(t)$ is a periodic complex function with frequency ${\Gamma}.$
Thus, if the frequencies satisfy
\begin{equation}
\label{eq:balance}
- n\,\Gamma +\Omega_2+\Omega_3 = {\deltaTb}
\end{equation}
for  some $n \in \mathbb{Z},$ then the amplitude $b_4(t)$ grows linearly in time without bound. This establishes the instability.

Crucially, if one re-scales all amplitudes $b_j$ by the same factor, then $\Gamma$ and $\Omega_j$ also re-scale by that factor. Therefore for any initial condition one can ``tune'' the initial amplitudes by a scaling factor in order to satisfy Eq.~(\ref{eq:balance}). This is a critical balance: a nonlinear frequency equates a linear frequency. It is worth emphasising that if $\deltaTb = 0$, such tuning is not possible. Therefore this mechanism favours energy transfers towards non-resonant triads.\\

\noindent \textbf{Triggering the nonlinear instability.} The system of equations (\ref{eq:ODEs}) can be reduced at most to four degrees of freedom, so numerical solutions are needed. A physically relevant Hasegawa-Mima plasma-wave example takes $\deltaSb = 0$, $\deltaTb = -\frac{8}{9},$ $S_1 = 1$, $S_2 = 9$, $S_3 = -8$, $T_1 = -1$, $T_2 = \frac{8}{3}$ and $T_3 = -\frac{9}{5}$.
A generic set of initial amplitudes is $b_1(0) = 0.007772 \,A$, $b_2(0) = 0.0385822\,A$, $b_3(0) = - 0.0358876 \,\mathrm{i} \,A$ and $b_4(0) = 0$, where $A$ is the so-called scale factor, a real constant to be fine-tuned in order to trigger the instability.
The instability (\ref{eq:balance}) is exact when $\epsilon = 0$ and approximate for finite values of $\epsilon,$ but it survives as ``persistent'' unstable periodic orbits \cite{Darboux1878,Fenichel1971}. Eq.~(\ref{eq:balance}) leads to predicted unstable $A$-values:
\begin{equation}
\label{Aeqn}
A_n=({0.3068\,n - 0.0796})^{-1},
\end{equation}
where $n \in \mathbb{Z}\,.$ The case $A_n<0$  is omitted here for simplicity.
Positive $A_n$ are obtained only for $n>0$.  If $|n|$ is too large, the initial push from Eq.(\ref{eq:solb4INST}) will be too small so in practice, only the cases $n=1, 2$ are relevant.  The predicted unstable $A$-values are:  $A_{1}=4.40$ and $A_{2}=1.87$.

\begin{figure}
\begin{center}
\includegraphics[width=0.4\textwidth]{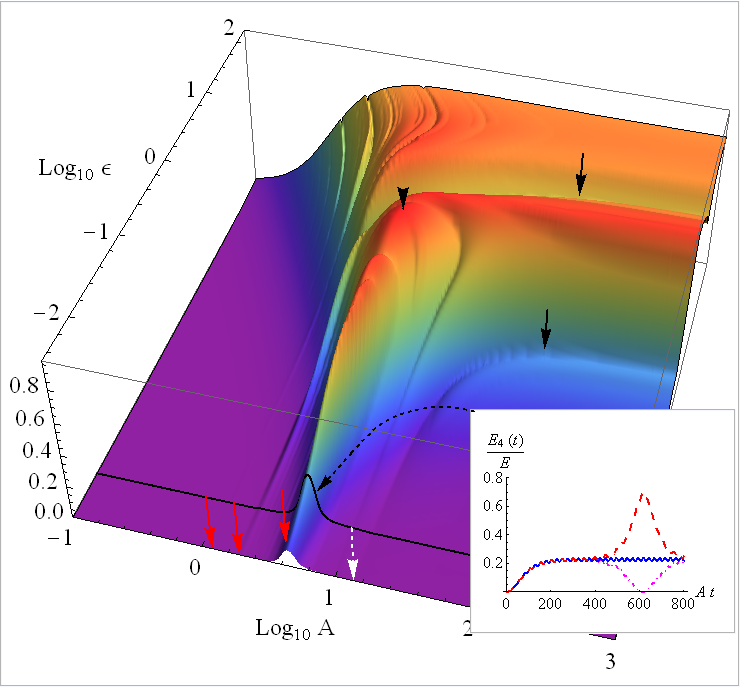}
\includegraphics[width=0.075\textwidth]{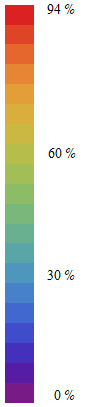}
\caption{\label{efficiency3D} (Colour online) Efficiency of energy transfer to $b_4$ as a function of $(\epsilon, A)$.
Red solid arrows: from left to right, predicted resonances $A_{3},A_{2},A_{1}$ at $\epsilon=0.$
White dashed arrow: sub-harmonic $A_{\frac{1}{2}}$-channel at $\epsilon=0.$
Top arrow without tail: global peak efficiency of $94\%$ at $A=7.36, \epsilon=0.97.$
Black solid arrows: periodic orbits at $A=100, \epsilon=0.881$ and $A=100, \epsilon=12.5.$ \textbf{Inset:} Taking $\epsilon=0.01$, $b_4$'s relative contribution to $E$ for $A=A_{*}\equiv 5.320174$ (solid curve), $A=A_{*} + 10^{-5} A_{*}$ (dot-dash curve) and $A=A_{*} - 10^{-5} A_{*}$ (dashed curve). }
\end{center}
\end{figure}

A really conclusive study of system (\ref{eq:ODEs}) requires the exploration of the transfer efficiency towards mode $b_4$ as a function of the parameters $\epsilon$ and $A.$ To wit, the relative contribution of the mode $b_4$ to the positive-definite quadratic invariant $E \equiv \frac{72}{5}|b_1(t)|^2+\frac{9}{5}|b_3(t)|^2+\frac{8}{3}\,\epsilon \, |b_4(t)|^2$ is evaluated, by simulating numerically a family of systems (\ref{eq:ODEs}) parameterised by $\epsilon$ and scale factor $A$, for a simulation time of the order $100/A$ (this ensures that each simulation covers about 5 nonlinear periods $\frac{2\,\pi}{\Gamma}$). For each run, the efficiency is defined as the maximum relative transfer of invariant $E$ towards mode $b_4$ during the simulation time. Fig.~\ref{efficiency3D} shows the profile of transfer efficiency as a function of $A$ and $\epsilon.$

The whole landscape in Fig.~\ref{efficiency3D} is covered by a network of persistent unstable periodic orbits \cite{Darboux1878,Fenichel1971} (periodic modulo precession frequencies), evidenced as ``channels'' and ``ridges'' of the transfer efficiency profile that originate at $(\epsilon,A) = \left(0, A_{\frac{p}{q}}\right)$ (Eq.~\eqref{Aeqn}) with $p, q \in \mathbb{N}$ coprime. The three main ridges, marked with solid arrows at the left side of the plot, correspond to the predicted values $A=A_{1}, A_{2}, A_{3}.$ Channels, always of positive curvature, correspond to $A=A_{\frac{p}{q}},$ with $q > 1$ ($A_{\frac{1}{2}}$ is marked with a dashed arrow). It is evident that any periodic solution of system \eqref{eq:ODEs} is unstable, since the original system is volume-preserving.

The role of periodic orbits at the main ($A_1$) ridge of transfer efficiency is exemplified by the case $\epsilon=0.01,$ and three very close initial conditions, obtained by changing slightly the scaling factor: $A=A_* \equiv 5.320174$ (asymptotically a periodic orbit) and $A=A_\pm \equiv (1 \pm 10^{-5}) A_*.$ For these parameter values, the inset in Fig.~\ref{efficiency3D} shows the corresponding three time evolutions of the relative contribution of  mode $b_4$ to quadratic invariant $E,$ for a simulation time of $40$ nonlinear periods. Notably, when $A = A_{\pm}$ the solution hits the unstable manifold of the periodic orbit (excursion time $\approx 15$ nonlinear periods). Lyapunov exponents, computed using the QR method \cite{Dieci1997}, have a ratio $-2:-1:3.$ In general near a ridge or channel in Fig.~\ref{efficiency3D}, the unstable manifolds of the periodic orbits have an eigendirection towards large $|b_4|$ values. By hitting the unstable manifolds, the system is allowed to explore higher values of $|b_4|$ and also lower ones, depending on which side of the unstable manifold the system is.

Near the main peak in Fig.~\ref{efficiency3D} there are several intersections between ridges and channels, corresponding to a series of period bifurcations leading to sub-harmonic periodic solutions. A crucial result is that all periodic orbits found numerically at the channels and ridges have a period close to $ \frac{\tau_0\,q}{A}$, where $q$ is the denominator that classifies the channel ($q > 1$, sub-harmonic) or ridge ($q = 1$) and $\tau_0$ is close to the \emph{fixed} period ($\approx 23.0$ for this simulation) of the original isolated source triad,
computed analytically at $\epsilon=0, A=1.$  This result is justified theoretically by exploiting the dilatation symmetry of Eqs.~\eqref{eq:ODEs}. To illustrate this, the $q=1$ periodic orbit at the main ridge ($\epsilon=0.01, A = A_{*}$, linked to the inset in Fig.~\ref{efficiency3D}) has period $\tau_0 \approx 24.4,$ and the $q=2$ periodic orbit at the sub-harmonic $A_{\frac{1}{2}}$-channel ($\epsilon = 0.881, A=100$, solid arrow above the inset in Fig.~\ref{efficiency3D}) has period $\tau_0 \approx 20.3.$ In this sub-harmonic case the unstable manifold that gives rise to strong transfers has a nontrivial winding number in the full phase space (see supplemental material). The same happens with the unstable manifold at $\epsilon = 12.5, A =100$ (top-right solid arrow in Fig.~\ref{efficiency3D}).

The importance of this result is that at any point $(\epsilon, A)$ not on a periodic orbit, the typical period of the oscillation can be found explicitly and is close to the period of the nearest strongest ridge or channel.\\

\noindent \textbf{Network of transfers: Cascades.} As $\epsilon$ grows from $0.01$ to $1$ and beyond to $100,$ persistent ridges and channels of transfer efficiency in Fig.~\ref{efficiency3D}
proceed towards large-$A$
values (the usual high-nonlinearity regime). There, instabilities are triggered by deforming $\epsilon,$ not $A.$ But changing
$\epsilon$ leads to physically different realisable triad connections. Although one needs an extra parameter apart from $\epsilon$ in order
to sample all possible triad connections, the enormous variability of the transfer efficiency with respect to $\epsilon$ demonstrates that
different triads in a given cluster will behave differently only because the interaction coefficients are different, even with the same type of initial conditions.

\begin{figure}
\begin{center}
\includegraphics[width=0.45\textwidth]{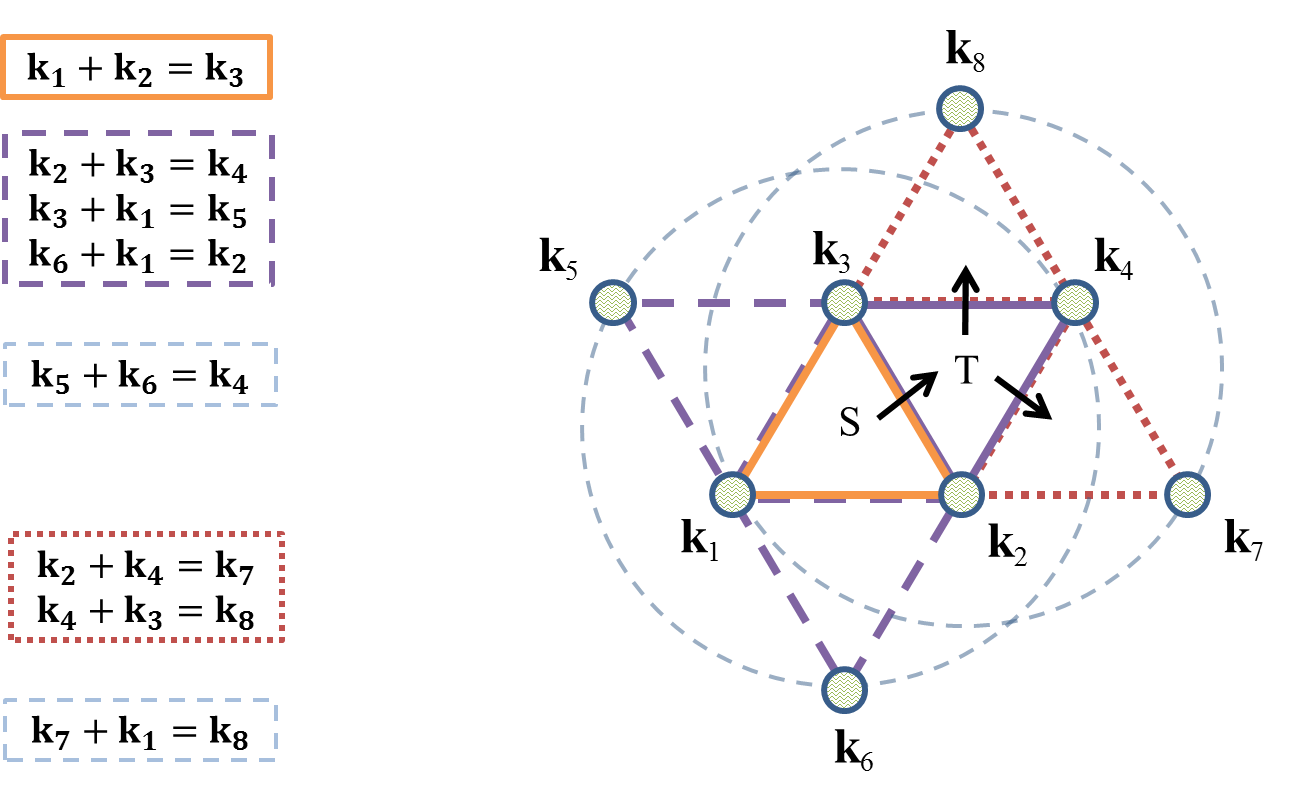}
\caption{\label{8triads} (Colour online) Source triad $\{{\bt k}_1,{\bt k}_2,{\bt k}_3\}$ with all first-layer target modes ${\bt k}_4$, ${\bt k}_5, {\bt k}_6$ and some second-layer target modes such as ${\bt k}_7$ and ${\bt k}_8$. Circles denote algebraically-dependent triads. In total, $8$ triads are shown. At any level of cluster truncation the system has two Manley-Rowe type of invariants. Arrows indicate hypothetical directions of efficient transfers of these invariants out of the source triad $S.$}
\end{center}
\end{figure}

The source triad $\{\bt{k}_1, \bt{k}_2, \bt{k}_3\}$ is part of an extended cluster of connected triads as shown in Fig.~\ref{8triads}. The turbulent cascading process consists of
5 stages:\\
\textbf{1.} Out of the geometrically available target modes $b_4, b_5, b_6,$ one
will receive the highest transfer according to the previous analysis of instability landscapes. Let us say that $b_4$ receives most
of this transfer with the $(\epsilon,A)$ parameters near a ridge. Then the four modes $b_1, b_2, b_3, b_4$ will behave quasi-periodically
with period $\frac{\tau_0\,q}{A},$ where $q \in \mathbb{N}$ and $\tau_0$ is the typical source-triad nonlinear period at $A=1.$\\
\textbf{2.} The modes $b_4, b_5, b_6$ form an algebraically dependent triad: $\bt{k}_5 + \bt{k}_6 = \bt{k}_4$
follows from the defining relations (see Fig.~\ref{8triads}). This new triad is very relevant physically, since it gives rise to one extra term in the evolution equations for $b_4, b_5, b_6.$ If, for this new triad, 
the mode $b_4$ is unstable in the usual sense~\cite{Craik1988}, then the decay instability will transfer energy to the modes $b_5, b_6,$ leading to a multi-periodic collective oscillation. If  $b_4$ is stable then the energy will not go towards modes $b_5, b_6.$ \\
\textbf{3.} The next stage is a new layer of geometrically available target modes $b_7, b_8$ stemming from $b_4$ (etc. for
$b_5$ and $b_6$), each mode being part of new triads connected
to the previous triads via two common modes. The question of which mode gets the energy is a repetition of
Eqs.~\eqref{eq:ODEs}--\eqref{eq:balance} with, say, $b_7$ as target mode and $b_4, b_2$ as source modes with period $\frac{\tau_0\,q}{A}$.
Therefore, a new efficiency landscape will be generated and one of the two new modes $b_7, b_8$ will have a higher transfer efficiency,
corresponding to the ridge/channel in the efficiency landscape associated with the period $\frac{\tau_0\,q\,q'}{A},$ where $q' \in \mathbb{N}.$\\
\textbf{4.} The modes $b_1, b_7, b_8$ form another algebraically dependent triad:  $\bt{k}_1+\bt{k}_7 =\bt{k}_8.$ Again the decay instability may
redistribute the energy in that new triad.\\
 \textbf{5.} Iterating the above processes leads to a cascade of efficient energy transfers. The cascade path in wave-vector space is formed by concatenation of two-common-mode connections between triads, conserving two invariants at any stage \cite{Harper2012}. At each stage, a new layer of target modes is produced but the selected mode(s) will depend on the particular transfer efficiency profile which in turn depends on the frequency mismatches, interaction coefficients and initial conditions. Thus, the detailed energy-transfer path depends on these quantities and the most efficient path is through connected triads with roughly similar values of frequency mismatch $\delta$, so that efficiency landscape for each triad is near the corresponding main ridge. This assertion is backed up with recent work on percolation of non-resonant triads, where the size of the cluster of connected triads shows a transition at a critical value of the allowed frequency mismatch of the triads \cite{Harris2013}. We have confirmed the above steps 1-5 in direct PDE numerical simulations. The detailed calculations will be presented elsewhere.\\

\noindent \textbf{Concluding remarks.} The robust energy transfer mechanism found in this paper provides a quantitative understanding of turbulent cascades in nonlinear wave systems of finite size, in terms of periodic orbits/unstable manifolds. Selection of cascading paths along clusters of connected triads depends on the relative strengths of the interactions and initial amplitudes. It favours energy exchanges towards non-resonant triads due to a critical balance between linear and nonlinear frequencies. The effect is likely to be found in experiments/observations.

\section{ACKNOWLEDGMENTS}
We thank C. Connaughton, F. Dias, J. Dudley, P. Lynch and S. Nazarenko for useful discussions.
We are deeply grateful to the organisers of the ``Thematic Program on the
Mathematics of Oceans'' at the Fields Institute, Toronto, who provided support
and use of facilities for this research. Additional support for this work was
provided by UCD Seed Funding project SF652,  IRC Fellowship ``The nonlinear
evolution of phases and energy cascades in discrete wave turbulence'' and ERC-2011-AdG 290562-MULTIWAVE.

\bibliography{bibliography}
\end{document}